\documentclass[aps,prd,showpacs,twocolumn,groupedaddress]{revtex4}

\usepackage{graphicx}

\begin{document}

\title{Double Extensive Air Shower Induced by Ultra-High Energy Cosmic
  Tau-Neutrino}

\author{M. M. Guzzo}
\email[]{guzzo@ifi.unicamp.br}
\author{C. A. Moura Jr.}
\email[]{moura@ifi.unicamp.br}
\affiliation{Instituto de F\'{\i}sica Gleb Wataghin - UNICAMP\\
             13083-970 Campinas SP, Brazil}

\date{\today}

\begin{abstract}
We investigate the possibility of detecting ultra-high energy cosmic
tau-neutrinos by means of a process involving a double extensive air
shower, the so-called Double-Bang Phenomenon. In this process a
primary tau-neutrino interacts  with an atmospheric quark creating a
first hadronic shower and a tau-lepton,  which subsequently decays
creating a second cascade.  The number of  these events strongly
depends on the flux of tau-neutrinos arriving at the Earth's
atmosphere and can be used to test  some theoretical models related to
the production of ultra-high energy tau-neutrinos. We estimate the
potential of the  fluorescence detector of the Pierre Auger
Observatory to observe Double-Bang events. We conclude that  for
tau-neutrinos with energies ranging from $O(0.1)$~EeV to $O(10)$~EeV
the number of detected events vary from  hundreds in a year to only
few events in hundreds of years.
\end{abstract}


\maketitle

\section{\label{intro}Introduction}
It is believed that ultra-high energy cosmic neutrinos may
play an important role to explain the origin of cosmic rays with
energies beyond the GZK limit of few times $10^{19}~$eV~\cite{G,ZK},
once that neutrinos hardly interact with cosmic microwave background
or intergalactic magnetic fields, keeping therefore its original
energy and direction of propagation. Even if they have masses or
magnetic moments, or travel distances of the order of the visible
universe, those characteristics do not change very much.
Possible sources of these ultra-high energy neutrinos, like Active
Galactic Nuclei and Gamma Ray Bursts, are typically located at
thousands of Mpc~\cite{HSaltzberg,Halzen}. Considering that neutrinos come
from pions produced via the process $\gamma + p \rightarrow N + \pi$
\cite{Halzen}, that there is an additional $\nu_e$ flux due to
escaping neutrons and that about 10\% of the neutrino flux is due to
proton-proton ($pp$) interactions, the proportionality of different
neutrino flavors result:
$\nu_e:\nu_{\mu}:\nu_{\tau}=0.6:1.0:<0.01$~\cite{learned}.
Nevertheless, observations of solar~\cite{solar} and
atmospheric~\cite{atmospheric} neutrinos present compelling evidence
of neutrino flavor oscillations. Such oscillations have been
independently confirmed by terrestrial
experiments. KamLAND~\cite{kamland} observed $\bar{\nu}_e$
disappearance confirming (assuming CPT invariance) what has been seen
in solar neutrino detections and K2K~\cite{ktokplb,ktokprl} observed
$\nu_{\mu}/\bar{\nu}_{\mu}$ conversion compatible with what has been
detected in atmospheric neutrino observations.

In order to understand these experimental results by means of neutrino
oscillations, two scales of mass squared differences and large mixing
angles have to be invoked. For solar and KamLAND observations, $\Delta
m^2_{\odot}\sim7\times10^{-5}~$eV$^2$ and
$\sin^22\theta_{\odot}\sim0.8$. And for atmospheric neutrino and K2K,
$|\Delta m^2_{atm}|\sim3\times10^{-3}~$eV$^2$ and
$\sin^22\theta_{atm}\sim1$.
Moreover LSND experiment~\cite{lsnd} may have observed
$\bar{\nu}_{\mu}\rightarrow\bar{\nu}_e$ transition which can be also
explained by neutrino oscillations with a large mass scale, $|\Delta
m^2_{LSND}|\sim(0.5-2.0)~$eV$^2$. Such results will soon be checked by
MiniBooNE~\cite{mini}.
These scales require four neutrino oscillation framework (or three, if
LSND results will not be confirmed by MiniBooNE experiment) which
imply, for ultra-high energies of the order 1~EeV or higher,
oscillation lengths much smaller than typical distances from the
sources of ultra-high energy neutrinos.
Consequently when neutrino flavor oscillations are taken into
consideration the flavor proportion will be modified to
$\nu_e:\nu_{\mu}:\nu_{\tau}\sim1:1:1$. Therefore one expects a
considerable number of tau-neutrinos arriving at the Earth.

In this paper we investigate the possibility of detecting ultra-high
energy cosmic tau-neutrinos by means of a process in which a
double Extensive Air Shower (EAS) is identified, the so-called
Double-Bang (DB) Phenomenon.
In that kind of event a tau-neutrino interact  with a quark via
charged current  creating one cascade of hadronic particles and a
lepton tau which subsequently decays producing a second cascade.
DB Phenomenon was first proposed for detectors where the neutrino
energy should be around $1~$PeV~\cite{learned}.  It does not happen
with neutrinos different from tau. The electron generated by an
electron-neutrino immediately interacts after being created and the
muon generated by a muon-neutrino, on the other hand, travel a much
longer distance than the size of the detector before interacting or
even decaying. So we do not have DB events from them.

In order to identify a DB Phenomenon we have to look for  two
Extensive Air Showers (EAS) in the same direction of propagation
inside the field of view (f.o.v.) of the detector, i.e., in the
physical space around the detector in which an event can be triggered.

In the Pierre Auger Observatory~\cite{auger}, a hybrid detection
technique will be used to make a detailed study of cosmic rays at
energies mainly around 10~EeV and beyond. The two techniques consist
of an array of detectors spread on the ground (the ground array
detector also called surface detector)
and an optical detector used to probe longitudinal development of
EAS by recording the fluorescence light emitted by
the excited nitrogen molecules of the Earth's atmosphere.
It has been shown that the Auger Observatory can detect atmospheric
near-horizontal air showers generated by neutrinos with the surface
detector~\cite{PaZa,CCPZ,Wilczy,Ave,Letessier,athar}. We concentrate here on
a different approach to study the possibility of detecting events
induced by  ultra-high energy tau-neutrinos,  the DB events, using the
fluorescence detector of that observatory.

We conclude that the features of the Pierre Auger Fluorescence Detector
favor the observation of DB events with tau-neutrino energies
varying from  $O(0.1$~EeV) to $O(10$~EeV), despite the low efficiency
of the fluorescence detector at energies smaller than 1~EeV. We
estimate the number of DB events observed in the fluorescence detector
varying  from hundreds in a year to few events in hundreds of years
depending mainly on the primary tau-neutrino flux.

This paper is organized in the following way: Section~\ref{DBAuger}
has a brief introduction to the DB Phenomenon. Section~\ref{sim} shows
some results of the DB events simulated and Section~\ref{nbev}
describe how we calculate the number of events in the Pierre Auger
Fluorescence Detector. Section~\ref{resdis} has the number
of events calculated for different models of ultra-high energy
neutrino flux and discuss how could we take some physical information
from that. The conclusions are in Section~\ref{conc}, in which we
discuss also the background events.

\section{\label{DBAuger}The Ultra-High Energy Double-Bang and the Auger
  Observatory}
Studying the characteristics of the fluorescence detector, such as its
efficiency and f.o.v. and the characteristics of the DB
events generated by ultra-high energy tau-neutrinos, one can estimate
the rate of that kind of event expected in the Auger Observatory.

\begin{figure}
\resizebox{0.48\textwidth}{!}{
\includegraphics{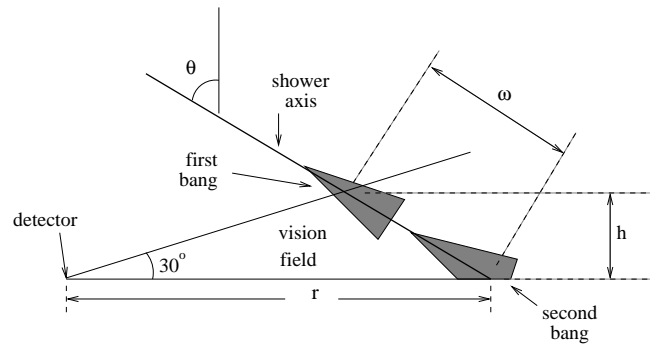} }
\caption{\label{esqdb}A schematic view of a Double-Bang and the f.o.v.
  of the Pierre Auger Fluorescence Detector. See text for details.}
\end{figure}

Fig.~\ref{esqdb} shows a schematic view of an ultra-high energy DB
with the detector position and the time integrated development of the
two showers, one created by the high-energy tau-neutrino interacting
with a nucleon in the atmosphere and the other created by the decay of
the tau generated in the first interaction of the tau-neutrino. The
f.o.v. of the fluorescence detector will be comprehended between
angles near the horizontal ($\sim2^o$) and $30^o$, and a radius $r$ of
approximately 30~km. The maximal height from where the DB can be
triggered by the fluorescence detector is h and $\omega$ is its
projection in the DB propagation axis. The zenith angle is represented
by $\theta$.

We considered only showers moving away from the detector since, in the
opposite case,  a large amount of \v{C}erenkov light arrives together
with the fluorescence light, spoiling a precise data
analysis~\cite{bellido}.

The total amount of light emitted by the first cascade is related to
the energy transfered to the quark at the moment of the first
tau-neutrino  interaction, which we will define as $E_1$. The neutrino
energy $E_{\nu}$ is the sum of the tau energy $E_{\tau}$ and $E_1$,
i. e., $E_{\nu}=E_1+E_{\tau}$. For charged current interactions above
0.1~EeV, approximately 20\% of the neutrino energy is transfered to
the quark~\cite{walker}. The second cascade, resulting from the tau
decay, carries an energy $E_2$ of approximately $2/3~E_{\tau}$ and is
visible when the tau decay is hadronic, which happens with  a
branching ratio of around 64\%~\cite{pdg}.

Therefore, very roughly, we have $\left<E_1\right>\sim1/5E_{\nu}$ and
$\left<E_2\right>\sim2/3\left<E_{\tau}\right>\approx8/15E_{\nu}$ and
the relation between $E_1$ and $E_2$ is given by:
$E_2/E_1=\frac{8}{15}E_{\nu}/\frac{1}{5}E_{\nu}\approx2.67$. The
distance traveled by the tau before decaying in laboratory frame is
$L={\gamma}ct,$ where $\gamma=E_\tau/m_\tau$ and $t$ is the mean life
time of the tau, that has an error of approximately 0.4\%~\cite{pdg}.

\subsection{\label{sim}Simulations}
In order to infer the possibility of observing DB events in atmosphere
and detecting them in the Pierre Auger Fluorescence Detector,  we
first simulate the longitudinal development of the showers.  For
simplicity, we numerically simulate DB events using protons as primary
particles. We create two separated showers using the relations of
energy ($E_2/E_1$) and distance ($L$) between the two EAS of a DB in
the way presented in the previous section.  In fact, it has been
argued that it is difficult to distinguish if a ultra-high energy
EAS was created by a proton or a
neutrino~\cite{cronin,jain}, at least when the cross section is
extrapolated from the Standard Model. An important  difference is the
probability of interaction in the atmosphere, that can be $10^{-5}$
smaller for a vertical neutrino than for a proton. There are also some
works that study the differences between the longitudinal development
of EAS generated by protons, heavier nuclei and different neutrino
flavors~\cite{jain,Ambrosio}. In reference~\cite{Ambrosio} they use
CORSIKA+Herwig Monte Carlo simulations to have electron and
muon-neutrinos as primary particles, but not tau-neutrinos. So the
simulations we have made to study DB events still are a good
approximation. We discuss how to distinguish events that could
masquerade tau-neutrino induced DB events in the conclusions.

Two different approaches can be taken to evaluate the longitudinal
development of a DB. At first we used Gaisser-Hillas
parameterization~\cite{gh}. This method, nevertheless is not reliable
to determine the depth of the shower maximum and the point of first
interaction, specially for arrival angles larger than $60^o$. At this
point, the CORSIKA simulation seems to be a good
approach~\cite{moura}. Here we use its version 6.00.

\begin{figure*}
\resizebox{0.6\textwidth}{!}{
\includegraphics{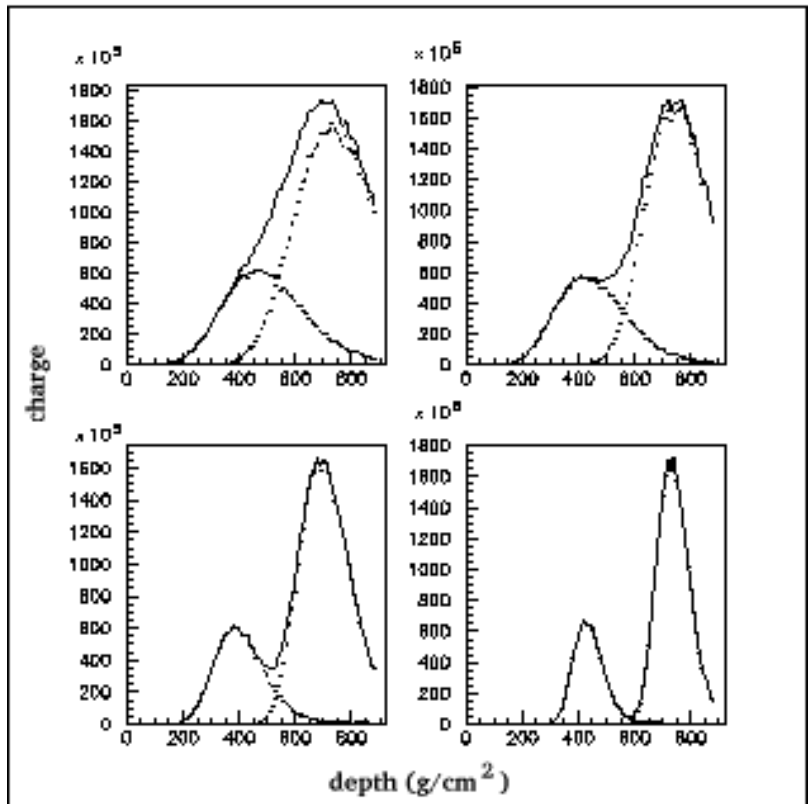} }
\caption{\label{sim1}Number of charged particles as a
function of atmospheric depth in g/cm$^2$ simulated for incident
angles of $45^o$, $55^o$, $65^o$ and $75^o$ from left to right and up
to down. Inferior lines represent the first and second EAS, and the
upper line is the sum when the two EAS are superimposed. The energy
of the primary neutrino is 0.5 EeV.}
\end{figure*}

\begin{figure*}
\resizebox{0.6\textwidth}{!}{
\includegraphics{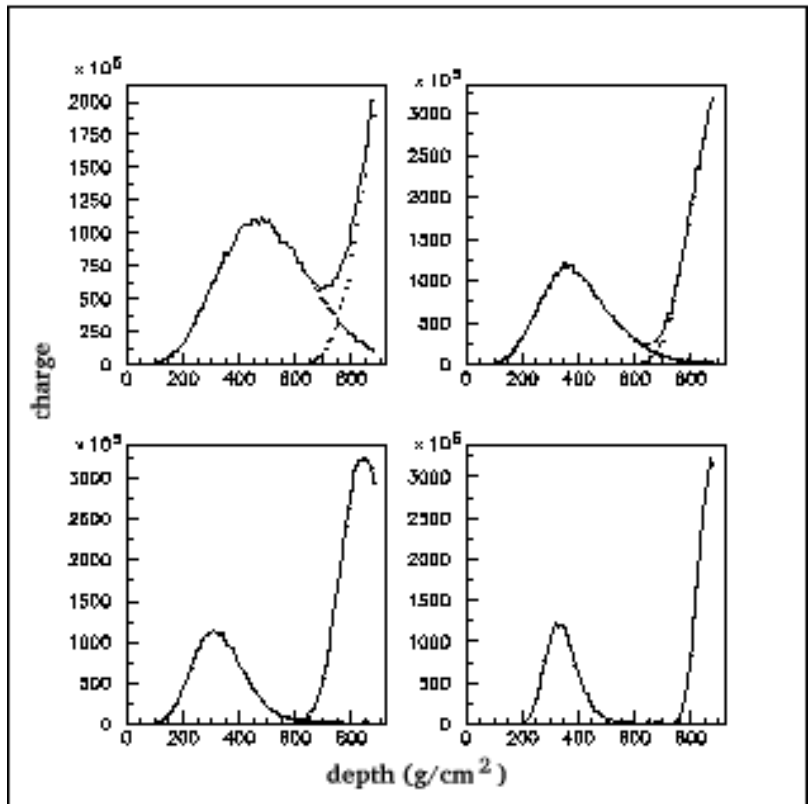} }
\caption{\label{sim2}Same as Fig.~\ref{sim1}, but the energy of the
  primary neutrino is assumed 1 EeV.}
\end{figure*}

Fig.~\ref{sim1} and Fig.~\ref{sim2} present the main results of the
simulations with the CORSIKA program. They show the longitudinal
development of the DB for different incident angles chosen among the
simulated events. Table~\ref{tab1} and Table~\ref{tab2} show the
parameters used in the simulations. In these tables, the points of
first and second interactions are chosen fixed parameters. The
distance between the first and second 
interactions, as we pointed out above, is in accordance with the tau mean
decay length ($L$) in the laboratory frame. Because of the small cross
section of the neutrino, the probability of interaction in the top of
the atmosphere is about the same for any point.

\begin{table}
\caption{\label{tab1}Parameters used in the numerical
simulations with the CORSIKA program and primary neutrino energy of
0.5 EeV. For a visualization of the geometrical parameters $\theta$, h and
$\omega$, see Fig.~\ref{esqdb}. The units are km except where it is
pointed.}
\begin{ruledtabular}
\begin{tabular}{ccccccc}
$\theta$\footnote{Zenith angle measuring the shower incident axis
    inclination} &
{\it 1st. int.}\footnote{Altitude of the first interaction in the
    atmosphere} &
$L\cos\theta$\footnote{Projection in the vertical axis of the distance
    the tau runs before it decays} &
{\it 2nd. int.}\footnote{Altitude where the tau decays} &
h\footnote{Maximum altitude from where the fluorescence light of the DB can be
    detected, for $r=30$~km (see Fig.~\ref{esqdb})} &
h (g/cm$^2$)\footnote{Same as h, but in units of g/cm$^2$, also called
    atmospheric depth} &
$\omega$\footnote{Projection of h in the DB propagation axis} \\
\hline
$45^o$ & 24.3 & 14.3 & 10  & 11  & 231 & 15.6 \\
$55^o$ & 19.6 & 11.6 & 8   & 9.5 & 291 & 16.6 \\
$65^o$ & 15.5 &  8.5 & 7   & 7.7 & 377 & 18.2 \\
$75^o$ & 10.2 &  5.2 & 5   & 5.5 & 515 & 21.3 \\
\end{tabular}
\end{ruledtabular}
\end{table}

\begin{table}
\caption{\label{tab2}Same as Table~\ref{tab1}, where the primary neutrino
  energy is taken to be 1~EeV.}
\begin{ruledtabular}
\begin{tabular}{ccccccc}
$\theta$ & {\it 1st. int.} & $L\cos\theta$ & {\it 2nd. int.} & h &
h (g/cm$^2$) & $\omega$ \\
\hline
$45^o$ & 33.6 & 28.6 & 5   & 11  & 231 & 15.6 \\
$55^o$ & 28.2 & 23.2 & 5   & 9.5 & 291 & 16.6 \\
$65^o$ & 22.1 & 17.1 & 5   & 7.7 & 377 & 18.2 \\
$75^o$ & 13.5 & 10.5 & 3   & 5.5 & 515 & 21.3 \\
\end{tabular}
\end{ruledtabular}
\end{table}

\subsection{\label{nbev}The number of events}
To calculate the possible number of events in the Pierre Auger
Fluorescence Detector, first consider for simplicity one detector with
a site seeing of 360$^o$. Then we can write the equation:
\begin{equation}\label{eq1}
N_{events}=\int P_{det}(E,r,\theta,\varphi)~\Phi_{\nu}(E)~dE~dA~dT~d\Omega
\end{equation}
where,
\begin{equation}\label{eq:pdet}
P_{det}(E,r,\theta,\varphi) = P_{int}(E,\theta) \times
                              P_{trig}(E,r,\theta,\varphi)
\end{equation}
is the probability of detection of a DB in the Auger Observatory,
given by the product of the probability $P_{int}(E,\theta)$ of the
tau-neutrino to interact in the atmosphere and the probability
$P_{trig}(E,r,\theta,\varphi)$ of the DB to be triggered by the fluorescence
detector.
$\Phi_{\nu}(E)$ is the flux of high-energy neutrinos depending on the
model of the extragalactic source of high-energy cosmic rays. $E$,
$A$, $T$ and $\Omega$ are the energy of the incident neutrino, area
over the f.o.v. of the detector, time of data acquisition and
solid angle around the detector, respectively.

The interaction probability is given approximately by:
\begin{equation}\label{eq:pint}
P_{int}(E,\theta) = \langle \sigma_{{\nu}N}(E) \rangle \times \langle
  N_T(\chi) \rangle
\end{equation}
where $\langle\sigma_{{\nu}N}(E)\rangle$ is the average cross section
of the neutrino-nucleon interaction and $\langle N_{T}(\chi)\rangle$,
the average total number of nucleons per squared centimeter at the
interaction point in the atmosphere. $N_{T}(\chi)=2N_A\chi(\theta)$,
where $N_A$ is the Avogadro's number and $\chi(\theta)$ is the
atmospheric slant depth at the neutrino-nucleon interaction point.

\begin{figure}
\rotatebox{270}{ \resizebox{0.42\textwidth}{!}{
\includegraphics{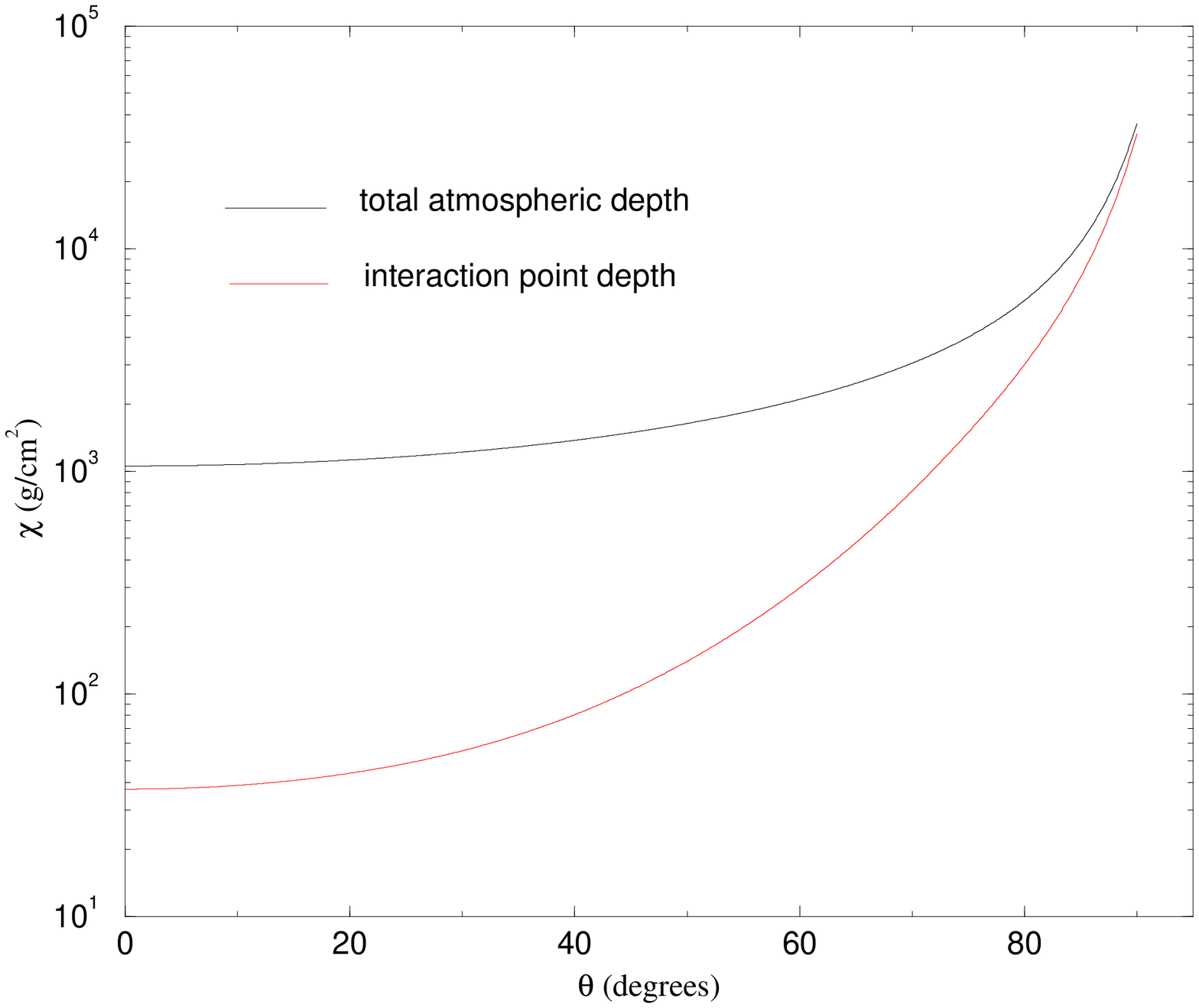} }
}
\caption{\label{densxangul}Atmospheric depth as a
  function of the zenith angle. The inferior line represents the
  interaction point depth for $l=30$~km.}
\end{figure}

Considering the Earth's curvature, the slant depth can be
approximately written as:
\begin{equation}\label{eq:prof}
\chi(\theta)=\int_\lambda\rho(H=lcos\theta+\frac{(lsin\theta)^2}{2R})d\lambda
\end{equation}
where $\lambda$ is the path along the arrival direction from infinity
until the interaction point in the atmosphere, $\rho$ is the
atmospheric density, $H$ the vertical height, $l$ is the distance
between the interaction point and the point toward the particle goes
through on Earth (the slant height), and $\theta$, the zenith
angle. The atmospheric depth as a function of the zenith angle is
shown in Fig.~\ref{densxangul}.

As the cross section of neutrinos with ultra-high energies is unknown,
usually one adopts the extrapolation of parton distribution functions
and Standard Model parameters far beyond the reach of experimental
data. In this way, one can estimate a value for the cross section of
the neutrino-nucleon interaction of about $10^{-32}$~cm$^2$, for
energies around 1~EeV. Some
authors say that this extrapolation gives a neutrino-nucleon cross
section that is too high~\cite{weiler} but others use models that
increase this same cross section to typical hadronic cross section
values~\cite{jain}. In this work we use the extrapolation of the
Standard Model cross section.

The trigger probability is given by:
\begin{equation}\label{eq:ptrig}
P_{trig}(E,r,\theta,\varphi)=\Upsilon\times P_{had}\times
                     P_{db}(E,r,\theta,\varphi)\times\Sigma(E,r)
\end{equation}
where $\Upsilon$ is the fraction of the time the fluorescence detector
will work ($\Upsilon=0.1$ because the fluorescence detector can only
operate in clear moonless nights), $P_{had}$ is the hadronic branching
ratio of tau decay ($P_{had}=0.64$), $P_{db}(E,r,\theta,\varphi)$ is the
probability of the DB to be seen by the detector and $\Sigma(E,r)$ is
the efficiency of the detector.

We define the probability of the DB to be seen by the fluorescence
detector as:
\begin{eqnarray}\label{eq:pdb}
P_{db}(E,r,\theta,\varphi) &=& 
                   \frac{\omega(r,\theta,\varphi)}{L(E)}\mbox{, if }
                         \omega(r,\theta,\varphi) \le L(E) \nonumber \\
                           &=& 1\mbox{, otherwise}
\end{eqnarray}
where, as we have seen in Section~\ref{DBAuger}, $L(E)$ is the
distance traveled by the tau in laboratory frame and
$\omega(r,\theta,\varphi)$, as can be seen in the Fig.~\ref{esqdb}, is the
size of the shower axis inside the f.o.v. of the detector. We made a rough
approximation to account only for the showers moving away from the
detector, so that $\int\omega(r,\theta,\varphi)d\varphi \sim
\pi\omega(r,\theta)$. Then we measure $\omega(r,\theta)$
where the vertical plane containing the shower axis passes
through the center of the fluorescence detector.

The efficiency $\Sigma(E,r)$ is the convolution of the energy
efficiency $\Sigma'(E)$ and the efficiency depending on distance
$\Sigma''(r)$. As the fluorescence detector has no measurement of
$\Sigma'(E)$ yet, but only expectations based on simulations, we
constructed one function based on simulations that grows
logarithmically from 0 to 1 in the interval of neutrino incident
energy 0.3~$<E~({\rm EeV})~<$~88~\cite{barbosa}. $\Sigma'(E)=0$ if
the energy is lower than the energies in this range and $\Sigma'(E)=1$
if the energy is higher than that.
$\Sigma''(r)$ is a Gaussian distribution centered at the point 15~km
far from the detector, but for simplicity in the calculation we used a
step function 20~km long centered at the same point. It does not make a
great difference in the final result.

\section{\label{resdis}Results and Discussion}
Because the Auger Observatory Fluorescence Detector will be
constituted of four detectors with a site seeing of 180$^o$,
we have to multiply the result of Eq.~\ref{eq1} by 2. Then using
Eq.~\ref{eq1} we calculated the number of events which
can be seen in Tables~\ref{tab5} and~\ref{tab6} for different models
of ultra-high energy cosmic ray flux and in different energy intervals.
\begin{table}
\caption{\label{tab5}Number of events in the
  fluorescence detector during a period of three years, calculated in
  different regions of the energy spectrum and for different models of
  neutrino flux. TD-92 stands for the model in reference~\cite{bhat};
  TD-96 for the reference~\cite{sigl}; AGN-95J for~\cite{karl}; WB
  for~\cite{wb} and MPR for~\cite{mpr}.}
\begin{ruledtabular}
\begin{tabular}{lllll}
Models ($p$) &
$N_1$\footnote{$0.20<E($EeV~$)<0.63$} &
$N_2$\footnote{$0.63<E($EeV~$)<2.00$} &
$N_3$\footnote{$2.00<E($EeV~$)<6.30$} &
$N_4$\footnote{$6.30<E($EeV~$)<200$} \\
\hline
TD-92 (0)   & 260  & 450    & 180   & 70   \\ 
TD-92 (0.5) & 2.8  & 5.7    & 2.8   & 1.5  \\
MPR         & 2.4  & 3.6    & 1.1   & 0.3  \\
TD-92 (1)   & 0.14 & 0.37   & 0.25  & 0.2  \\ 
AGN - 95J   & 0.15 & 0.22   & 0.07  & 0.02 \\
TD-92 (1.5) & 0.06 & 0.15   & 0.1   & 0.08 \\
WB          & 0.05 & 0.08   & 0.02  & 0.01 \\
TD-96 (1)   & $2.9\times10^{-5}$    & $3.1\times10^{-5}$    &
$6.1\times10^{-6}$    & $1.0\times10^{-6}$ \\
\end{tabular}
\end{ruledtabular}
\end{table}

From Table~\ref{tab5} one can learn which is the energy interval which is 
relevant to detect DB events in the Auger fluorescence detector.
It is, approximately, 0.63~$<E~({\rm EeV})~<$~2. The neutrino flux used
here for the model AGN-95J is an approximation that takes roughly the
average between the models A and B in
reference~\cite{karl}. Table~\ref{tab6} shows the expected number of
DB events in three years according to the ultra-high energy cosmic
rays source models, when a Standard Model extrapolation for the
$\sigma_{{\nu}N}$ is used.
%
\begin{table}
\caption{\label{tab6}Tau neutrino fluxes at the Earth according to the
  corresponding total number of events in a period of three years.}
\begin{ruledtabular}
\begin{tabular}{llc}
Models ($p$) & $\Phi_{\nu}$ (GeV$^{-1}$m$^{-2}$s$^{-1}$sr$^{-1}$) &
$N_{events}$(3yrs)$^{-1}$ \\
\hline
TD-92 (0)   & $2.31\times10^{-2}E^{-1.77}$ & $9.6\times10^{2}$ \\
TD-92 (0.5) & $5.64\times10^{-6}E^{-1.58}$ & $1.3\times10^{1}$ \\
MPR         & $2.15\times10^{-2}E^{-2}$    & $7.4\times10^{0}$ \\
TD-92 (1)   & $1.07\times10^{-9}E^{-1.3}$  & $9.6\times10^{-1}$ \\
AGN - 95J   & $1.34\times10^{-3}E^{-2}$    & $4.6\times10^{-1}$ \\
TD-92 (1.5) & $4.24\times10^{-10}E^{-1.3}$ & $3.8\times10^{-1}$ \\
WB          & $4.83\times10^{-4}E^{-2}$    & $1.7\times10^{-1}$ \\
TD-96 (1)   & $7.51\times10^{-4}E^{-2.4}$  & $6.7\times10^{-5}$ \\
\end{tabular}
\end{ruledtabular}
\end{table}

Based on the simulations shown in Section~\ref{sim}, one can have an
idea of the longitudinal development of the DB as a function of the
incident angle and energy of the primary neutrino. It is the
convolution of the terms $P_{db}(E,r,\theta,\varphi)$ and
$\Sigma(E,r)$ in Eq.~\ref{eq:ptrig} that describes the restriction in
the energy interval observed in the simulations and
Table~\ref{tab5}. For relatively low energies the efficiency
$\Sigma(E,r)$ of the detector is also low and, as can be seen in
Eq.~\ref{eq:pdb}, for energies higher than 2~EeV the factor
$P_{db}(E,r,\theta,\varphi)$ is too small.

\begin{figure*}
\resizebox{0.6\textwidth}{!}{
\includegraphics{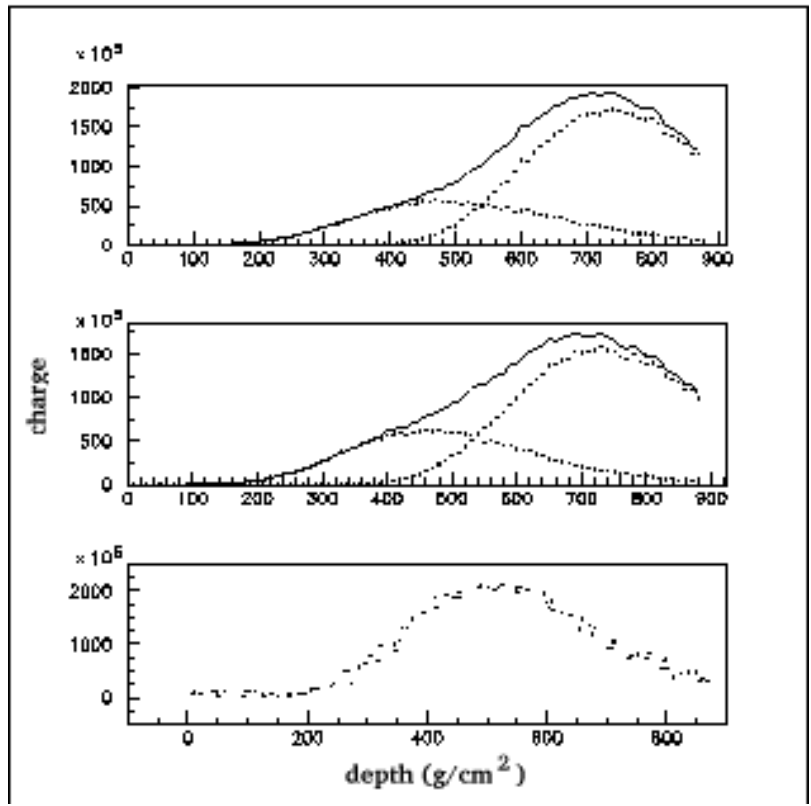} }
\caption{\label{comp}Number of charged particles as a function of
atmospheric depth in g/cm$^2$ simulated for an incident angle of
$45^o$ and depth of the first interaction equal to 24.3km $\approx$
30g/cm$^2$. The first two graphics from the top were simulated using
the CORSIKA program versions 5.62 (which is valid for zenith angles
smaller than 60$^o$) and 6.00 (which is valid even for zenith angles
larger than 60$^o$) respectively, with the energy summed of the two
EAS equal to 0.5~EeV. Inferior lines represent the first and second
EAS, and the upper line is the sum when the two EAS are
superimposed. The graphic in the bottom is an ordinary shower
simulated with the CORSIKA version 6.00 and energy of 0.5 EeV.}
\end{figure*}

There is also a restriction for the neutrino incident zenith
angle. For zenith angles smaller than 55$^o$, because the lower
atmospheric density, the development of the shower is slower and there
is a superposition of the two EAS. The DB still can be triggered but
in a more sophisticated way by comparing the longitudinal development
of the EAS. Fig.~\ref{comp} shows events with incident angles equal to
$45^o$. In the simulations on the top of Fig.~\ref{comp} we can observe the
two overlapped showers such that they look like a single ordinary
EAS. Even in that case one can see that it is possible to identify a
DB because both the maximum of the charged particles number and the
longitudinal development of the shower are not compatible with a
``single-bang'' of the same energy, which is shown in the lowest plot of
Fig.~\ref{comp}.  For zenith angles larger then 75$^o$, since h (see
Fig.~\ref{esqdb}) is too low, it is difficult to detect the first EAS.

We simulated sixty DB events for each different energy and angle that
we present. The results have a root mean square deviation for the
maximum of the showers of approximately 30 g/cm$^2$. As the maximum of
the DB (top and central plots) in Fig.~\ref{comp} is around 700
g/cm$^2$ and the maximum of the ordinary EAS (lowest plot) is around
500 g/cm$^2$, it is easy to differentiate between the depth of the
maximum of a DB and an ordinary EAS in the simulations we made. If a
higher energy proton will masquerade a lower energy DB depends on the
accuracy of the detector to determine the energy of the primary
particle. Because of the mean life time of the tau, only relatively
low energy DB will be superimposed looking like an ordinary EAS. So if
you detect an ordinary EAS profile of relatively high energy
($E>10$~EeV), that cannot be considered a DB event.

To study the relation between the cross section and the potential of
the Pierre Auger Fluorescence Detector to detect DB events we wrote
the neutrino cross section as $\sigma_{{\nu}N}=A\times\sigma_{SM}$,
where $A$ is a free parameter which depends on the model and
$\sigma_{SM}\sim7.8\times10^{-36}$~cm$^2\times E^{0.363}$ is the
Standard Model extrapolation for $\sigma_{{\nu}N}$ which have 10\%
accuracy within the energy range
$10^{-2}<E($EeV$)<10^3$ when compared with the results of the
CTEQ4-DIS parton distributions~\cite{gandhi}.

To account for models that gives a cross section different from the
extrapolation of the Standard Model cross section we just vary $A$. It
is a naive approximation but gives an idea of how the number of
events can increase or decrease if the cross section is not
standard. For instance, if the cross section is of order 10 times
higher then the standard in the energy range $0.20<E($EeV$)<200$,
then even the WB upper bound for the ultra-high energy cosmic ray flux
predict a rate of more than 1 event in 3 years in the Pierre Auger
Fluorescence Detector.
\begin{figure}
\rotatebox{270}{
\resizebox{0.40\textwidth}{!}{
\includegraphics{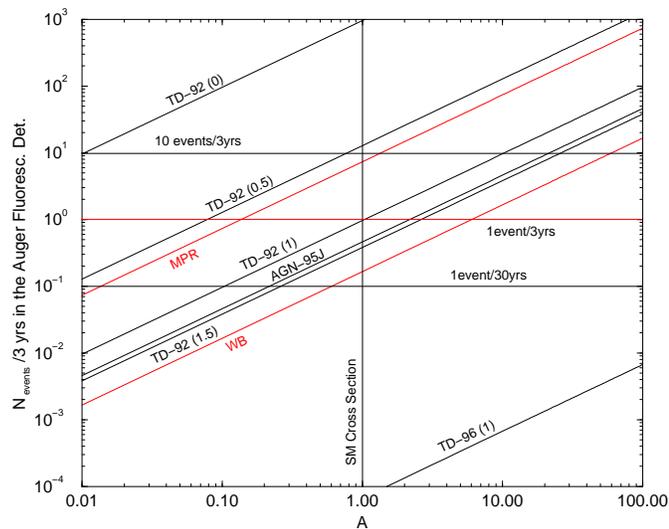}
}
}
\caption{\label{fvsc}Number of events according to
  the parameter $A$ of the cross section for various models. $A=1.00$
  corresponds to the extrapolation of the Standard Model cross
  section. See text for details.}
\end{figure}

The horizontal lines in Fig.~\ref{fvsc} are number of events
chosen for comparison (1 event in 3 years is
considered the lower limit for the rate of events in the Auger
Observatory). The vertical line shows the extrapolated Standard Model
cross section. The other lines show the number of events for different
models of ultra-high energy cosmic ray flux depending on $A$.
For instance, we can see from this figure that assuming the Standard
Model extrapolated cross section, if the Observatory measure a number
of events around one in three years, models like TD-96(1), TD-92(0)
and TD-92(0.5) will be excluded, and the more compatible with the
measurements will be TD-92(1).

\section{\label{conc}Conclusion}
Taken into consideration neutrino oscillations, one expect that one
third of the high-energy neutrino flux arriving at the Earth should be
composed  of tau neutrinos. These neutrinos  can interact in the
Earth's atmosphere generating a double shower event named Double-Bang
Phenomenon.  Many recent works have studied the potential of the
Pierre Auger Observatory to detect horizontal air showers generated by
ultra-high energy neutrinos with the surface detector. Here we
specifically investigate the potential of the fluorescence detector of
that observatory to  observe DB events.

DB events have very particular characteristics in the Auger
Observatory. Different from the neutrino events in the surface
detector, DB events do not need to come from the very near-horizontal
angles. Despite the low probability of interaction in the atmosphere,
we can also have tau neutrinos creating DB events with incident angles
from approximately 55$^o$ to 75$^o$. DB events also have a lower
energy, around 1~EeV, different from the energies around 50~EeV and
beyond expected for an ordinary EAS generated by the highest energy
cosmic rays.  In the range of energy approximately between 0.6~EeV and
2~EeV a considerable part of the two EAS that characterize a DB can be
detected by the fluorescence detector and then we have a DB trigger.

The number of DB events depends on many parameters like arrival flux,
cross section, energy and incident direction of the neutrinos, and
efficiency of the detector. Models like MPR, TD-92(1) and AGN-95J
generate a number of events of around one in three years in the Pierre
Auger Fluorescence Detector. This is because the energy range where the
DB can be detected is very strict. For energies less then 0.6~EeV the
efficiency of the detector is too low and for energies greater then
2~EeV the two EAS are too separated.

Concerning the background, the probability for a proton to generate a
DB and masquerade the DB generated by a
neutrino depends on two possibilities: 1) that the primary proton
interaction generates some fragment that will give rise to a secondary
shower deep in the atmosphere with energy higher then the first. 2)
that another shower created by some independent particle interacts
deep in the atmosphere masquerading the second EAS of the DB.

In the possibility 1, the second EAS will be created by the decay or
interaction of the fragment deep in the atmosphere. We consider that
the primary proton looses roughly half of its energy to the secondary
particles that constitute the EAS, and so it is very hard that the
second EAS has more energy than the first one. Now, considering that
for this high-energies we have a cosmic ray flux of the order of 1
particle per km$^2$ per year and that the only particles that could
interact deeply in the atmosphere are neutrinos, generating the second
independent EAS near the detector, the chance that the proton and this
second independent neutrino come from the same solid angle direction
interacting in a time interval of the tau decay in the laboratory
frame of $\gamma t\approx 131\times \frac{E_\nu}{\mbox{\scriptsize
[EeV]}}~\mu$s is approximately (at most) 1 in 10$^8$, what exclude the
possibility 2. The direction of the two EAS can be identified
specially if two fluorescence detectors trigger the same DB event
(with only one detector, it must be difficult to know the direction of
the EAS in the plane that contains the EAS and the detector).

Based on this assumptions, $E_2/E_1$ can be a good parameter to
identify DB events if the error and the average in the energy measure
is within certain specific value. The error in the energy measured by
the fluorescence detector will depend mainly on the atmospheric
conditions but hardly will exceed 50\%. For a DB event the situation
is optimistic because the most important is the relation between the
energies of the two EAS and this error is smaller than the error of
the absolute energy of an ordinary EAS. We can make a conservative
estimation of the error in the average ratio $E_2/E_1$ considering the
error in the absolute energy of 50\%. This will give a relative error
to the energy ratio of 70\%. So, since $E_2/E_1\approx2.67$ in average
as deduced in Section~\ref{DBAuger}, then considering such an error we
find 95\% of the events such that the energy ratio $E_2/E_1 > 1$. Then
$E_2/E_1 > 1$ is a good parameter to identify DB events.

Despite the fact the DB Phenomenon can be very rare, it is very
important to be prepared for its detection, specially in case the
Pierre Auger ground array detect near-horizontal air showers which can
indicate a sign for electron and/or muon neutrinos.  Consequently
oscillations imply a considerable number of tau neutrinos too. With
such a motivation, the Auger Observatory could calibrate its trigger
to be more sensitive to energies around 1~EeV or increase the number
of fluorescence detectors.

The potential of the DB Phenomenon to acquire valuable information
both in particle and astrophysics is irrefutable. For instance, the
cross section and flux of the ultra-high energy neutrinos are
speculative and can be investigated with DB events.

\begin{acknowledgments}
We thank Carlos Escobar, Vitor de Souza, Henrique Barbosa, Ricardo
Sato and Walter Mello of the Auger collaboration in the Instituto de
F\'{\i}sica at UNICAMP for valuable help and comments on the present work. This
research was partially supported by ``Conselho Nacional de
Desenvolvimento Cient\'{\i}fico e Tecnol\'ogico - CNPq'' and
``Funda\c{c}\~ao de Amparo \`a Pesquisa do Estado de S\~ao Paulo -
FAPESP''
\end{acknowledgments}

\bibliography{prdf}

\end{document}